\documentclass[aps,pre,superscriptaddress,twocolumn,balancelastpage]{revtex4-1}

\usepackage[colorlinks,bookmarks=false,citecolor=blue,linkcolor=blue,urlcolor=blue]{hyperref}
\usepackage[all]{hypcap}   

\usepackage{physics}

\usepackage{amsmath,amssymb}
\usepackage{graphicx}
\graphicspath{{figures/}{/}}

\usepackage{verbatim}
\usepackage{color}
\usepackage{ulem}

\usepackage{placeins}  
\usepackage{flafter}     

\usepackage{color}

\newcommand{\ue}{\text{e}}
\newcommand{\ui}{\text{i}}

\newcommand{\HIDDEN}[1]{}

\begin{document}

\title{Characterizing quantum chaoticity of kicked spin chains}

\author{Tabea Herrmann}
\affiliation{TU Dresden,
Institute of Theoretical Physics and Center for Dynamics,
 01062 Dresden, Germany}

\author{Maximilian F. I. Kieler}
\affiliation{TU Dresden,
Institute of Theoretical Physics and Center for Dynamics,
 01062 Dresden, Germany}

\author{Arnd B\"acker}
\affiliation{TU Dresden,
Institute of Theoretical Physics and Center for Dynamics,
 01062 Dresden, Germany}

\date{\today}

\begin{abstract}

Quantum many-body systems are commonly considered as quantum chaotic
if their spectral statistics, such as the level spacing distribution,
agree with those of random matrix theory (RMT).
Using the example of the kicked Ising chain we demonstrate that even if both level spacing
distribution and eigenvector statistics agree well with random matrix predictions, the entanglement
entropy deviates from the expected RMT behavior, i.e.\ the Page curve.
To explain this observation we propose a new
quantity that is based on the effective Hamiltonian of the kicked system.
Specifically, we analyze the distribution of the strengths of the effective spin interactions and
compare them with analytical results that we obtain for circular ensembles.
Thereby we group the effective spin interactions corresponding to the number $k$ of spins which contribute
to the interaction.
By this the deviations of the entanglement entropy can be attributed to significantly different behavior of the $k$-spin
interactions compared with RMT.

\end{abstract}

\maketitle

\section{Introduction}\label{sec:introduction}

The properties of many-body systems are commonly characterized using
spectral statistics.
In particular the comparison with the results of random matrix theory (RMT)
offers many insights. RMT was originally introduced to
describe the spectra of complex nuclei \cite{GuhMueWei1998,Meh2004}.
Later it was conjectured that, even for single-particle systems with classical
chaotic limit, the spectral statistics follow those predicted by RMT \cite{BohGiaSch1984}.
This has been confirmed for many systems and
several types of spectral statistics
and theoretically explained using semi-classical methods
\cite{Ber1985,SieRic2001,MueHeuAltBraHaa2009}.
The relation between classical chaos and RMT spectral statistics
has been used to introduce
the concept of quantum chaos also for many-body systems
which often do not have a classical limit:
A many-body system is commonly called \textit{quantum chaotic}
if the level spacing distribution follows
the predictions from RMT
\cite{JenSha1985,HsuAng1993,MonPoiBelSir1993,JacShe1997,AviRicBer2002,San2004,PonPapHuvAba2015,KheLazMoeSon2016,AkiWalGutGuh2016,BorLueSchKnaBlo2017,LuiBar2017b,KosLjuPro2018}.
Moreover, it is often expected (and assumed),
that if the level spacing distribution follows RMT,
also other statistical properties agree with the RMT predictions
\cite{SanRig2010a,AleRig2014,LuiLafAle2015}.
In particular, it is expected that for quantum chaotic many-body systems
the eigenstate entanglement
follows the results predicted from
RMT \cite{Lub1978,Pag1993,Sen1996,KumPan2011}.

In this paper we demonstrate that, even if a many-body systems'
level-spacing statistics is in excellent agreement with the RMT prediction,
i.e.\ it qualifies as being quantum chaotic,
the entanglement can be significantly lower than expected.
This is illustrated using two parameter sets for the prototypical
example of the kicked Ising spin chain. While for one chain
all considered properties follows RMT, 
we find very small deviations in the eigenvector statistics
and significant deviations in the entropy for the other chain
when comparing with the predictions from the RMT.
To understand the reason for such differing behavior we
propose to analyze the effective
Hamiltonians of the chains. For this we compute the strengths
of the individual effective spin interactions and group them
with respect to the number $k$ of spins which contribute to the
interaction. The distributions of the strengths are compared
with the analytical RMT results.
The analysis for the two parameter sets of the kicked spin chain
reveal good agreement of the $k$-spin interactions
with RMT for the chains which shows entanglement as predicted
from RMT. In contrast, a significantly different behavior of the $k$-spin
interactions for the chain with smaller entanglement entropy is found.
Here the 2-spin interactions are more pronounced, while the effective
interactions reduce with increasing $k$.

This paper is organized as follows:
In Sec.~\ref{sec:stat_prop} the basic concepts, namely
level-spacing statistics, eigenvector statistics, and entanglement,
are recalled. In Sec.~\ref{subsec:stat_prop_spin_chain}
the kicked Ising chain is introduced and
the numerical results for the two parameter sets are discussed.
The effective spin interactions are introduced in
Sec.~\ref{sec:decomp_H_eff}, together with the random matrix result,
and compared with the results for the two spin chains.
Finally, a summary and outlook is given in Sec.~\ref{sec:summary_outlook}.

\section{Statistical properties of kicked spin chains}\label{sec:stat_prop}

In the following we focus  on quantum systems described by a
unitary time evolution operator $U$ acting on an $N$ dimensional Hilbert space,
\begin{align}
        U|\psi_n\rangle = \ue^{\ui\varphi_n}|\psi_n\rangle\;,
        \quad \text{with}\quad
        n=1,2,\dots,N\;.
\end{align}
The eigenstates $|\psi_n\rangle$ are assumed to be normalized and the
eigenphases fulfill $\varphi_n \in [-\pi,\pi)$.
Furthermore, we choose the phases
$\{\varphi_1, \varphi_2, \dots, \varphi_N\}$ to be ordered increasingly.
For a quantum chaotic system it is expected, that the statistical properties
of both the spectrum and eigenstates
follow the predictions from random matrix theory.
For systems without any antiunitary symmetry
the statistics are described by the results
for the circular unitary ensemble (CUE) while
in the presence of an antiunitary symmetry (e.g., time-reversal)
the circular orthogonal ensemble (COE) has to be used \cite{Meh2004}.
In Sec.~\ref{subsec:stat_prop_spin_chain} we investigate the
properties of the kicked Ising spin chain, which is time-reversal invariant,
therefore we restrict to the results for COE in the following sections.

\subsection{Level spacing distribution}\label{subsec:spacing}

One of the simplest spectral statistics is the
distribution of the spacings $s_n$ between consecutive levels.
For the eigenphases one gets
\begin{equation}
  s_n = \frac{N}{2\pi}\left(\varphi_{n+1}-\varphi_n\right),
\end{equation}
with $\varphi_{N+1} := \varphi_1 + 2\pi$.
Here the pre-factor provides the unfolding, leading to
a unit mean spacing.
In the limit $N\rightarrow \infty$
the COE result is well-described by the
Wigner distribution \cite{Wig1967,DieHaa1990}
\begin{align}
        P_{\text{COE}} (s)
                \approx \frac{\pi}{2} s
                \exp\left(-\frac{\pi}{4}s^2\right)\;.
        \label{eq:spacing_COE}
\end{align}
Closely related to the level spacing distribution are the ratio statistics
which are particularly useful when an analytical expression for
unfolding of the levels is not available.
The ratios $\tilde{r}_n$ are defined by \cite{OgaHus2007}
\begin{align}
        \tilde{r}_n
                = \frac{\text{min}(s_n, s_{n-1})}{\text{max}(s_n, s_{n-1})}
                = \text{min}\left(r_n, \frac{1}{r_n}\right)\;,
        \label{eq:definition_tilde_r}
\end{align}
where $r_n = s_n / s_{n-1}$.
In Ref.~\cite{AtaBogGirRou2013}
an analytical prediction for the distribution of $r$ has been derived,
from which one gets $\langle r\rangle_{\text{COE}} = 1.75$ and
$\langle \tilde{r}\rangle_{\text{COE}} = 4-2\sqrt{3} \approx 0.5359$
for the COE case.

\subsection{Eigenvector statistics}\label{subsec:evec}

The second statistical property we are interested in is the distribution
of the components of the eigenvectors.
An eigenstate $|\psi_n\rangle$ is represented in
some  orthonormal basis $\{|m\rangle\}_{m=1,\dots,N}$
by the coefficients
$c_m^{(n)} = \langle m|\psi_n\rangle$.
Due to the normalization one has
$\sum_m |c_m^{(n)}|^2 = 1$.
With these coefficients one defines
\begin{align}
        \eta_{nm} = N |c_m^{(n)}|^2\;,
\end{align}
where the prefactor ensures
that the mean of $\eta_{nm}$ is one.
The distribution $P(\eta)$ of $\eta_{nm}$
is an often used characteristics of the properties
of the eigenstates.
In case of eigenstates of a COE matrix one gets in the limit $N\rightarrow \infty$
the Porter-Thomas distribution \cite{PorTho1956,BroFloFreMelPanWon1981,HaaGnuKus2018}
\begin{align}
        P_{\text{COE}}(\eta)
                = \frac{\exp(-\eta/2)}{\sqrt{2\pi\eta}} \;\;.
        \label{eq:evec_stat_COE}
\end{align}

\subsection{Von Neumann entropy}\label{subsec:page}
The third property we discuss is the eigenstate entanglement of bipartite systems.
Here we achieve this bipartite structure by splitting the
original $N$ dimensional system into two subsystems
of dimension $N_1$ and $N_2$ with $N = N_1 N_2$.
To quantify the entanglement, we use the von Neumann entropy, which
is defined for a state $|\psi\rangle$, by
\begin{align}
        S = - \text{tr}(\rho_1 \ln(\rho_1))\;,
        \label{eq:von_Neumann_entropy}
\end{align}
with $\rho_1 = \text{tr}_2(\rho)$ being the reduced density matrix of
subsystem 1, resulting from tracing out the second
subsystem from the density matrix
$\rho = |\psi\rangle\hspace*{-0.1cm}\langle\psi|$.
Unentangled states can be written as product states
$|\psi\rangle = |\psi_1\rangle\otimes|\psi_2\rangle$ and
have zero entropy, while
maximally entangled states have
$S_{\text{max}} = \ln\left(N_1\right)$.
The (Haar)-averaged entropy of random states from the COE
is slightly reduced \cite{Pag1993},
\begin{align}
        S_{\text{COE}} = \ln\left(N_1\right)
                - \frac{N_1}{2 N_2}\;.
        \label{eq:entropy_COE}
\end{align}
this formula holds for $1 \ll N_1 \le N_2$.
In Ref.~\cite{Sen1996} an exact formula for the entropy of CUE
states is given, and in Ref.~\cite{KumPan2011} for the COE and CSE,
which are valid without any restrictions to $N_1$ and $N_2$.
They all reduce to the above stated result \eqref{eq:entropy_COE}
for $1 \ll N_1 \le N_2$.
However, for the dimensions considered in this paper Eq.~\eqref{eq:entropy_COE}
by Page is accurate enough.
The dependence of the von Neumann entropy on the size of the first
subsystem $N_1$ is often called Page curve.

\subsection{Kicked Ising spin chain}\label{subsec:stat_prop_spin_chain}

\begin{table}
        \begin{tabular}{|l|llll|}
        \hline
            & $L$  & $J$   &  $M$  & $\{\Theta_n\}$\\
        \hline
          A & 12   &  0.80 & 1.35  &$\{7,7,7,7,8,8, 8,8,8,8,7,7\} \pi/32$\\
          B & 12   &  1.0  & 1.2   &$\{9,9,9,9,10,10, 10,10,10,10,9,9\} \pi/32$\\
          \hline
        \end{tabular}
        \caption{Parameter sets $A$ and $B$ for kicked spin chain.}
        \label{tab:parameter_A_B}
      \end{table}
\begin{figure*}
        \includegraphics{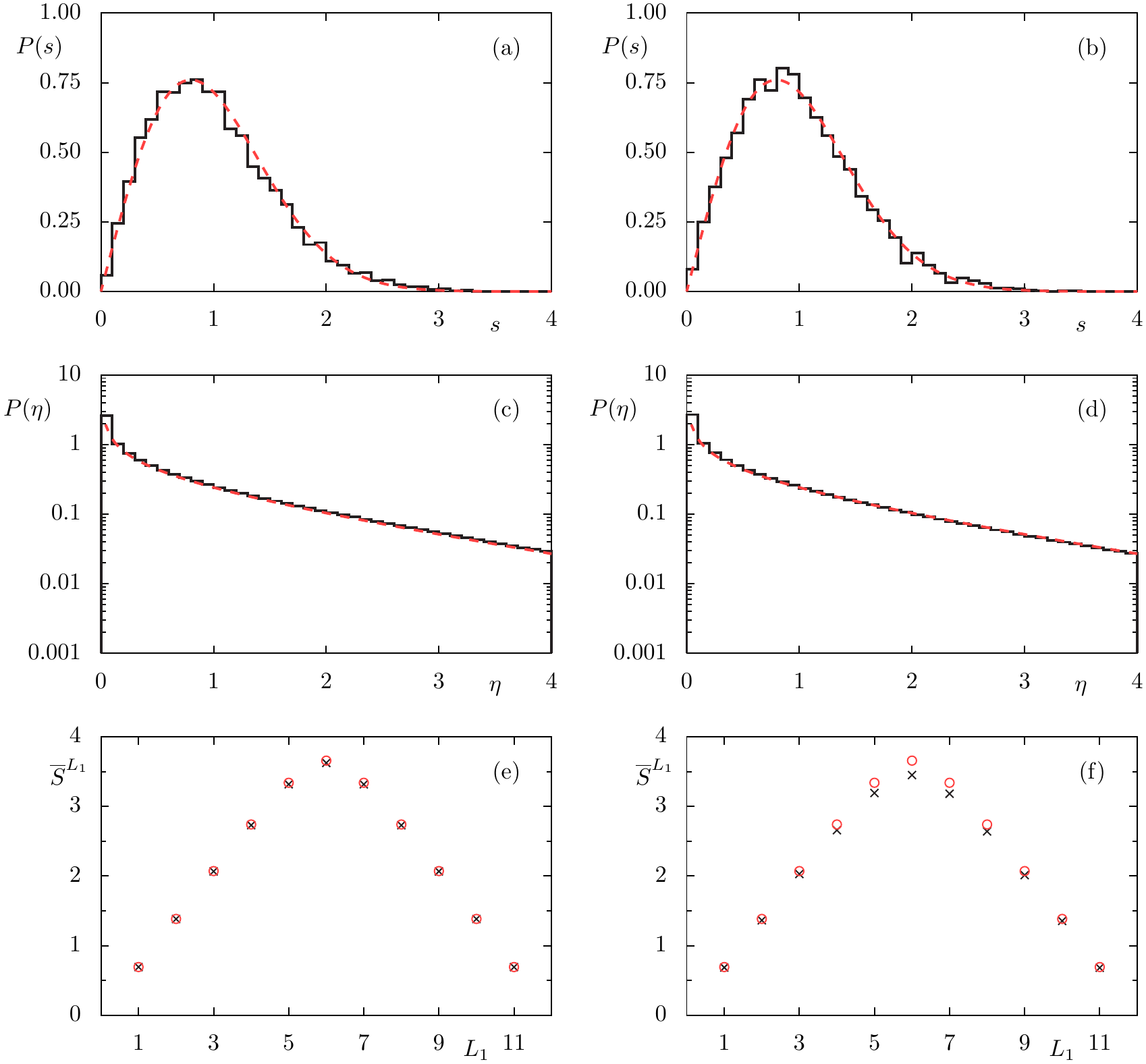}

        \caption{Level spacing distribution, eigenvector statistics, and von
        Neumann entropy for the kicked spin chain with $L=12$.
        Left column show results for parameter set A, right column
        for parameter set B, see table~\ref{tab:parameter_A_B}. Red
        dashed lines show COE results
        for level spacing distribution~\eqref{eq:spacing_COE} and
        eigenvector statistics~\eqref{eq:evec_stat_COE}. Red circles
        show predictions for random states~\eqref{eq:entropy_COE}.}
        \label{fig:fig_spacing_evec_page}
\end{figure*}

As an example of a many-body system we consider the
kicked Ising spin chain with open boundary conditions.
It is defined by the time dependent
Hamilton operator \cite{Pro2000,Pro2002,LakSub2005,MisLak2014,PinPro2007}
\begin{equation}
        H
                = H_{\text{free}} + H_{\text{kick}}
                \sum_{k=-\infty}^{\infty} \delta(t-k) \;,
        \label{eq:H_spin_chain}
\end{equation}
where
\begin{align}
        H_{\text{free}}
                =& J \sum_{n=1}^{L-1} \sigma^z_n \sigma^z_{n+1} \;,
        \label{eq:H_free}\\
        H_{\text{kick}}
                =& M \sum_{n=1}^{L}\left(\cos(\Theta_n)\sigma^x_n
                + \sin(\Theta_n)\sigma^z_n\right)
        \label{eq:H_kick}\;.
\end{align}
Here $\sigma^x_n$ and $\sigma^z_n$ are the standard Pauli
spin-matrices, see Eq.~\eqref{eq:pauli_matrices} below,
corresponding to the $n$th spin
and  $L$ is the number of spins in the chain so that the dimension of the
Hilbert space is $N=2^L$.
The free evolution part $H_{\text{free}}$ contains a nearest neighbor
coupling in the $z$-component with strength $J$.
The kicking part $H_{\text{kick}}$ displays a magnetic field of strength
$M$, which is periodically turned on and off represented by the sum over $\delta$
distributions.
{The magnetic field acts on all spins but the direction in which the kick is applied
for the $n$th spin depends on the angle of tilt $\Theta_n$ in the $x-z$ plane.}
The kicked Ising spin chain is time reversal invariant and has
a further symmetry which is called external reflection
or bitreversal symmetry \cite{PinPro2007, KarShaLak2007}. To avoid a
desymmetrization procedure
we break this symmetry by choosing different angles of tilt $\{\Theta_n\}$ for
the individual spins. {This allows us to do the following
numerics in the full Hilbert space of dimension $N$.}
We consider the following form of the time evolution operator
\begin{align}
        U
                =& \;\ue^{-\ui H_{\text{free}}/2}
                \ue^{-\ui H_{\text{kick}}}
                \ue^{-\ui H_{\text{free}}/2}\;,
        \label{eq_U_spin_chain}
\end{align}
for which the eigenstates have real coefficients in the computational basis.
For the numerical computations two slightly different parameter
sets for chain A and B are used, see table~\ref{tab:parameter_A_B}.

Figure~\ref{fig:fig_spacing_evec_page} shows the level-spacing distribution,
the eigenvector statistics, and the von Neumann entropy
for the two chains in comparison with the corresponding RMT results.
In Fig.~\ref{fig:fig_spacing_evec_page}(a) and (b) the level spacing
distribution is shown. We see that both chains
provide good agreement with the COE prediction.
Also, the results for the averaged
ratios $\langle r\rangle$ and $\langle \tilde{r}\rangle$ are
close to the results for COE: For chain~A we find
$\langle r\rangle_{\text{A}} = 1.735 \pm 0.052$ and
$\langle\tilde{r}\rangle_{\text{A}} = 0.5284 \pm 0.0039$
and for chain~B
$\langle{r}\rangle_{\text{B}}= 1.770 \pm 0.063$ and
$\langle\tilde{r}\rangle_{\text{B}} = 0.5342 \pm 0.0039$.
These agree within one standard deviation
with results for the COE stated in section~\ref{subsec:spacing}.
Thus, both chain~A and chain~B qualify as quantum chaotic.
Consequently, one could expect that also other more general
spectral statistics and in particular eigenstate statistics and
entanglement properties follow the RMT predictions.

The eigenvector statistics is shown in Fig.~\ref{fig:fig_spacing_evec_page}(c) and (d)
where the expansion coefficients of all eigenstates
of the time evolution operator \eqref{eq_U_spin_chain}
in the computational basis are used.
The histograms of $P(\eta)$ are shown in a semi-logarithmic representation.
Neither of the two chains shows a deviation from the Porter-Thomas
distribution.
Note that also for small values of $\eta$, very good agreement with the COE
is found, see App.~\ref{app:evec_ln}.

In Fig.~\ref{fig:fig_spacing_evec_page}(e) and (f)
the entanglement is shown for the two chains.
We compute for each eigenstate the
von Neumann entropy \eqref{eq:von_Neumann_entropy}
between two subchains, where subchain 1 contains the
first $L_1$ spins of the chain, i.e.\ $N_1=2^{L_1}$, and subchain 2 contains
the other $L_2=L-L_1$ spins, i.e.\ $N_2=2^{L-L_1}$.
From this the average entropy $\overline{S}^{L_1}$ of all eigenstates is obtained.
The corresponding Page curve, Eq.~\eqref{eq:entropy_COE},
is obtained using the corresponding dimensions $N_1$ and $N_2$
(with $N_1$ and $N_2$ exchanged if $N_1>N_2$.)

Chain~A agrees well with the Page curve, with some small visible deviations
around $L_1=6$.
In contrast, the entanglement of chain~B is significantly smaller.
As both chains qualify as quantum chaotic,
and also lead to good agreement with the RMT result
for the eigenvector statistics, this is unexpected.
It raises the question about the origin of this different
behavior for the two spin chains which we are going
to address in the following section.

\section{Effective spin interactions}\label{sec:decomp_H_eff}
The different behavior of the entanglement for chain~A and B in
Fig.~\ref{fig:fig_spacing_evec_page}, can be explained by differences
in the effective spin interactions.
To quantify these, we define an effective
Hamiltonian which contains these interactions.
Any unitary matrix $U$ can be written as $U = \ue^{-\ui H_{\text{eff}}}$,
where $H_{\text{eff}}$ is a Hermitian matrix
Thus, we define the effective Hamiltonian
corresponding to the unitary operator $U$ as
\begin{align}
        H_{\text{eff}} = \ui \ln(U)\;.
        \label{eq:H_eff_def}
\end{align}
We evaluate Eq.~\eqref{eq:H_eff_def} using the eigenvectors
and eigenphases defined in Sec.~\ref{subsec:spacing}, i.e.\
$ H_{\text{eff}} = - \sum_n \varphi_n \dyad{\psi_n}{\psi_n}$.
To analyze the effective spin interactions we
decompose the effective Hamiltonian into the spin interaction basis.
This basis is given by the tensor products
$\{S_1 \otimes \dots\otimes S_L\}$, where
$S_n \in \{\sigma_x,\sigma_y,\sigma_z,I\}$ for $n=1,\dots, L$,
with the Pauli matrices
\begin{align}
        \sigma_x={\begin{pmatrix}0&1\\
                1&0\end{pmatrix}}\;,\quad
        \sigma_y={\begin{pmatrix}0&-{\mathrm  {i}}
                \\
                {\mathrm  {i}}&0\end{pmatrix}}\;, \quad
        \sigma_z={\begin{pmatrix}1&0\\
                0&-1\end{pmatrix}}
        \label{eq:pauli_matrices}
\end{align}
and the identity matrix $I$.
With the Hilbert-Schmidt inner product (also called
Frobenius inner product for the
finite-dimenional case) one gets
an orthonormal basis.
Thus, we can write
\begin{align}
        H_{\text{eff}}
                = \sum
                \; C_{S_1, \dots, S_L}
                \left(S_1 \otimes \dots
                \otimes S_L\right)\;,
        \label{eq:H_eff_decompose}
\end{align}
where the sum runs over all combinations $(S_1 \otimes \dots \otimes S_L)$
and the $4^{L}$ real coefficients
$\{C_{S_1, \dots, S_L}\}$ are given by
\begin{align}
        C_{S_1, \dots, S_L} = \frac{1}{2^{L}}
        \text{Tr}\left(\left(S_1 \otimes
         \dots \otimes S_L \right)H_{\text{eff}}
         \right).
        \label{eq:def_coeff}
\end{align}
Each basis matrix is a Pauli string $(S_1 \otimes \dots \otimes S_L)$ and
represents a specific spin interaction,
e.g.\ for $L=4$ the basis matrix
$(\sigma_x\otimes \sigma_x \otimes I \otimes I)$ represents
the 2-spin interaction of the first and second spin in
the $x$ component
and $(\sigma_x\otimes \sigma_x \otimes \sigma_x \otimes I)$ the
3-spin interaction of first, second and third spin
in the $x$ component.

The coefficients $C_{S_1, \dots, S_L}$
describe the strength of the effective spin interactions.
A basis matrix containing $k$ Pauli matrices presents a $k$-spin interaction, and
we call $k$ the (interaction) order of the corresponding basis matrix.
Note that we only consider the number of non-trivial
factors in the basis matrix, i.e.\ we do not discuss effects of
locality or support of the basis matrix.
The distribution of the coefficients corresponding to the
same $k$-spin interaction order embody the effective strength
of this interaction order.
The only basis matrix with $k=0$ is the identity
matrix of dimension $2^L$.
Thus, the corresponding coefficient $C_{I, \dots, I}$
correspond to a global phase, and we set this
coefficient to zero by considering
an adjusted time evolution operator
$\tilde{U} = \ue^{\ui C_{I, \dots, I}} U$,
which fulfills $\tilde{C}_{I, \dots, I} = 0$.
In the following we omit the tilde, to simplify notation.

For systems with time reversal invariance there exists a basis
in which the eigenvectors are real \cite{BroFloFreMelPanWon1981,HaaGnuKus2018}.
This leads to a real effective Hamiltonians,
see~\eqref{eq:H_eff_eigen}, so that the coefficients
corresponding to imaginary basis matrices
are zero. Thus, the effective Hamiltonian is
fully described by the remaining $2^{L-1} (2^L+1)$ coefficients.
We only consider these coefficients in case of systems with an
antiunitary symmetry.

\subsection{Effective spin interactions for random matrices}\label{subsec:random_matrices}

\begin{figure}
        \includegraphics{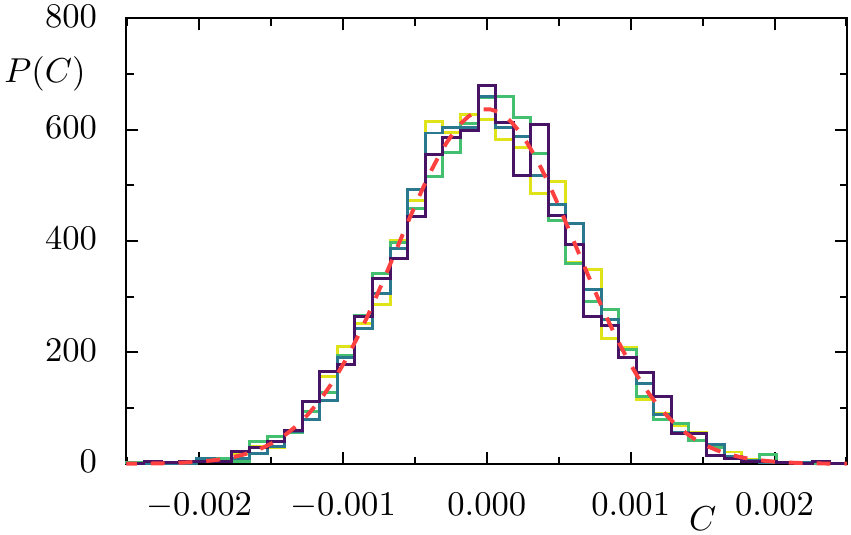}

        \caption{
        Distribution of the coefficients \eqref{eq:def_coeff}
        for COE matrices of size $N=2^{12}$ using 10 realizations.
        Coefficients for
        $k=2,4,7,12$ spin interactions are shown in dark to
        light colors.
        For each interaction order we use 500 coefficients
        of each realization.
        The red dashed line shows the prediction~\eqref{eq:coeff_distr_COE}
        for the  COE.}
        \label{fig:fig_coe_decompose}
\end{figure}

For random matrices from the COE and CUE the distribution
of the coefficients \eqref{eq:def_coeff} describing the
effective spin interactions can be obtained analytically.
For this we define the effective Hamiltonian corresponding to a COE matrix
$U^{\text{COE}}$ by
\begin{align}
        H_{\text{eff}}^{\text{COE}} = \ui \ln(U^{\text{COE}}).
        \label{eq:H_eff_COE}
\end{align}
In analogy to the kicked spin chains we evaluate Eq.~\eqref{eq:H_eff_COE}
using the eigenvectors and eigenphases of $U^{\text{COE}}$,
which are restricted to the interval $[-\pi,\pi)$.
Random matrices do not have any spin like structure, therefore
all interaction orders behave identical
and moreover all coefficients show the same distribution.
Using this we derive
in App.~\ref{app:random_matrix_result} that the distribution $P(C)$
of the coefficients for $N$ dimensional COE matrices is a normal distribution
\begin{align}
        P_{{\text{COE}}}(C)
        = \frac{1}
                {\sqrt{2\pi\sigma^2_{\text{COE}}}}
                \exp\left(-\frac{C^2}{2\sigma^2_{\text{COE}}}\right)\;,
        \label{eq:coeff_distr_COE}
\end{align}
with variance $\sigma^2_{\text{COE}}=2\pi^2 / (3 N^2)$.
Note that we find for CUE matrices the same distribution with
adapted variance $\sigma^2_{\text{CUE}}=\pi^2 / (3 N^2)$, see
App.~\ref{app:random_matrix_result}.

Figure~\ref{fig:fig_coe_decompose} compares the prediction
\eqref{eq:coeff_distr_COE}
with numerical data for $N=2^{12}$ dimensional COE matrices
using 10 realizations.
The distributions of the coefficients corresponding to
$k=2,4,7,12$ spin interactions are shown.
As expected, all interaction
orders behave similar and follow the predicted Normal distribution.

\subsection{Effective spin interactions for kicked spin chains}\label{subsec:decompose_spin_chains}

\begin{figure}
        \includegraphics{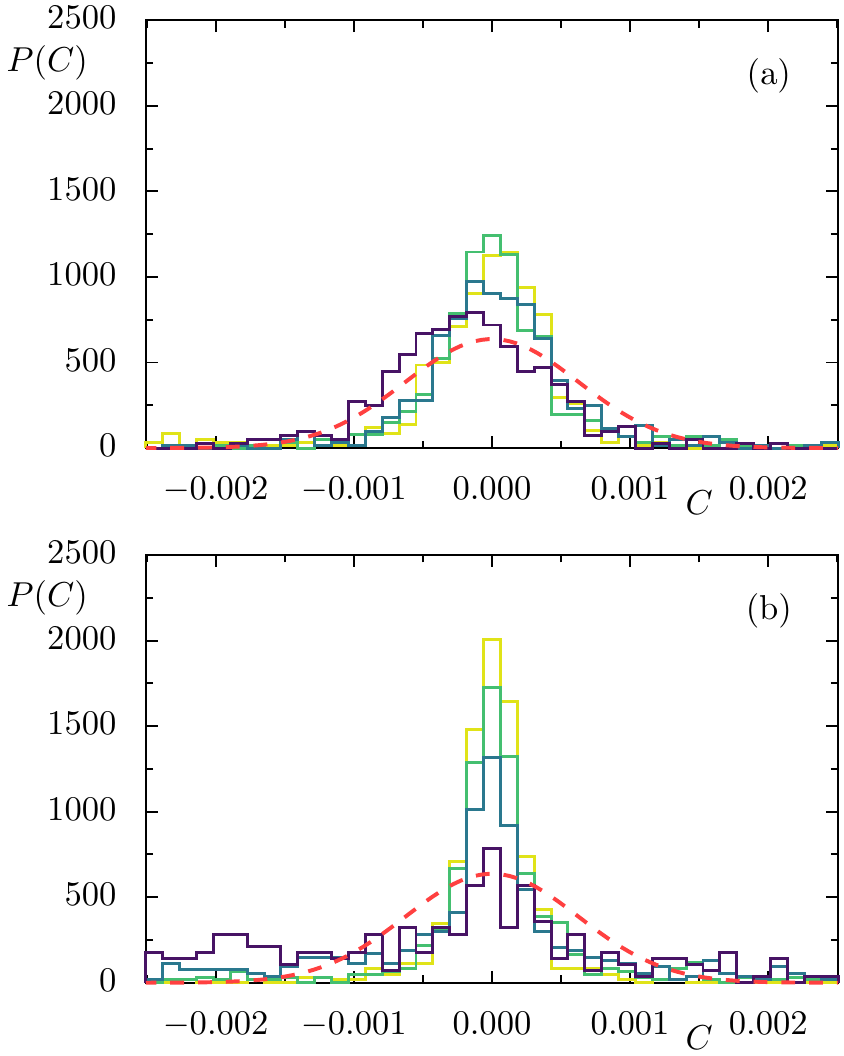}

        \caption{
        Distribution of the coefficients \eqref{eq:def_coeff}
        for kicked spin chains of length $L=12$.
        (a) shows result for parameter set A, (b)
        for parameter set B, see table \ref{tab:parameter_A_B}.
        Coefficients for
        $k=2,4,7,12$ spin interactions are shown in dark to
        light colors.
        For each interaction order
        500 coefficients are used.
        The red dashed line shows
        the prediction~\eqref{eq:coeff_distr_COE} for the COE.}
        \label{fig:fig_ktfim2_decompose}
\end{figure}

Now we compare the distribution of the effective spin
interactions for two spin chains with the random matrix prediction,
see Fig.~\ref{fig:fig_ktfim2_decompose}.
As for the COE case, we consider
$k=2,4,7,12$ spin interactions.\\
In Fig.~\ref{fig:fig_ktfim2_decompose}(a) we find for chain~A,
that the distributions for the different interaction
orders are all close to the predicted COE curve. The distribution for
the 2-spin interactions is slightly shifted to negative coefficients,
while the larger $k$-spin interactions are centered but a bit more
peaked than expected.
Nevertheless, we see in particular the same distribution for all $k>2$, which is
close to the RMT result and thus in line with the small deviations of the
entanglement shown in Fig.~\ref*{fig:fig_spacing_evec_page}.
In contrast,  Fig.~\ref{fig:fig_ktfim2_decompose}(b) for chain~B shows a
clearly different behavior of the effective interactions.
The distribution of the coefficients of the 2-spin interactions is broader
than the predicted normal distribution.
This means, that the 2-spin interactions are stronger than those of the COE.
For larger $k$ the distribution becomes more and
more peaked around zero, which means that these effective
spin interactions are too weak in chain~B.
The different behavior for the individual $k$-spin interactions illustrates
that we do not only find a different distribution of the interactions than expected for COE,
but a fundamentally different nature of spin interactions for chain~B.
As a consequence, this
does not allow the generation of
eigenstate entanglement predicted for COE states.
So the k-spin interactions, which get weaker with increasing k, explain
the smaller eigenstate entanglement, found
in Fig.~\ref*{fig:fig_spacing_evec_page} for chain~B.

\section{Summary, discussion, and outlook}\label{sec:summary_outlook}

In this paper we show, that if the level spacing
distribution of a system follows RMT,
the eigenstate entanglement does not necessarily show RMT results.
We demonstrate this by the example of two kicked spin chains, which both
qualify as quantum chaotic based on their level spacing distribution.
While for both chains the eigenvector statistics
follows the RMT prediction, the von Neumann entropy
differs from the expected Page curve.
For one chain we see the Page prediction nearly reached, but for the
other the eigenstate entanglement is significantly lower.
We can understand the observed entanglement
by analyzing the effective
$k$-spin interactions. In a random matrix situation the statitistics
of the interactions behaves identical for all $k$.
But for the second chain, with the
smaller eigenstate entanglement, we find that the interaction reduces with
increasing $k$ and thus behaves
significantly different from the COE. Therefore it does
not allow the creation of entanglement predicted by Page.
Interestingly, we also find for the first chain, which shows only very small
deviations from the Page curve, that the $k$-spin interaction statistics
is slightly different from the COE result.
Thus, the question arises if the deviations from RMT in the entanglement and
the effective spin interactions are typical for
quantum chaotic kicked spin chains.
Based on the numerical explorations to find good
parameter sets we have the impression that this is indeed
the case for the considered kicked Ising chain.
To properly decide this questions much longer chains would be
required where additionally the parameter dependence seems to become
less sensitive.
Moreover, it would be interesting to know if this different behavior
is also visible
in the long-range correlations of the levels or in a more detailed
analysis of the eigenvector components.

Note that deviations of the entanglement entropy from the expected
RMT results were also found for
the midspectrum eigenstates of autonomous spin chains
\cite{BeuAndHaq2015,VidRig2017,LiuCheBal2018,BeuBaeMoeHaq2018,LeBMalVidRig2019,BaeHaqKha2019}
including predictions and explanations for the deviations \cite{Hua2019,Hua2021,HaqClaKha2022}.

To describe the statistical properties of
eigenstates of many-body systems, several RMT models have been introduced,
which take the many body structure of the Hamiltonian into account.
One commonly used model are the embedded random matrices
\cite{FreWon1970,FreWon1971,BohFlo1971a,BohFlo1971b,BroFloFreMelPanWon1981,AsaBenRupWei2001,BenRupWei2001a,GomKarKotMolRelRet2011,Kot2014,BorIzrSanZel2016}
and power-law-banded random matrices \cite{KraMut1997,VarBra2000,EveMir2008,BogSie2018b}.
Another ansatz is to
use random 2-spin interactions to build a random Hamiltonian~\cite{KeaLinWel2015b}.
It would be helpful to have similar models for the unitary case.
One possible approach are Floquet random unitary
circuits \cite{KosLjuPro2018,ChaLucCha2018b,ChaLucCha2018a,ChaLucCha2019,MouPreHusCha2021}
which are an extension of random unitary circuits \cite{NahRuhVijHaa2017,vonRakPolSon2018}
to the Floquet setting and take the local structure of the time evolution operator into account
by applying repeatedly the same set of random gates which couple
neighboring spins. A different ansatz is
based on the effective Hamiltonian \eqref{eq:H_eff_def}
and its decomposition \eqref{eq:H_eff_decompose}
into the spin interaction basis.
If one can find the distribution for the coefficients
corresponding to a generic kicked spin chain
one can build random effective Hamiltonians $ H_{\text{eff}}^{\text{random}}$
and study $U = \exp\left({-\ui H_{\text{eff}}^{\text{random}}}\right)$
to investigate the properties of the corresponding time evolution operator.
This is an interesting task for the future.

\acknowledgments

We thank Felix Fritzsch, Masud Haque, Roland Ketzmerick, Ivan Khaymovich, and Shashi Srivastava for useful discussions.

Funded by the Deutsche Forschungsgemeinschaft (DFG, German Research Foundation) --  497038782.

\appendix

\section{Logarithmic eigenvector statistics}
\label{app:evec_ln}

\begin{figure}
        \includegraphics{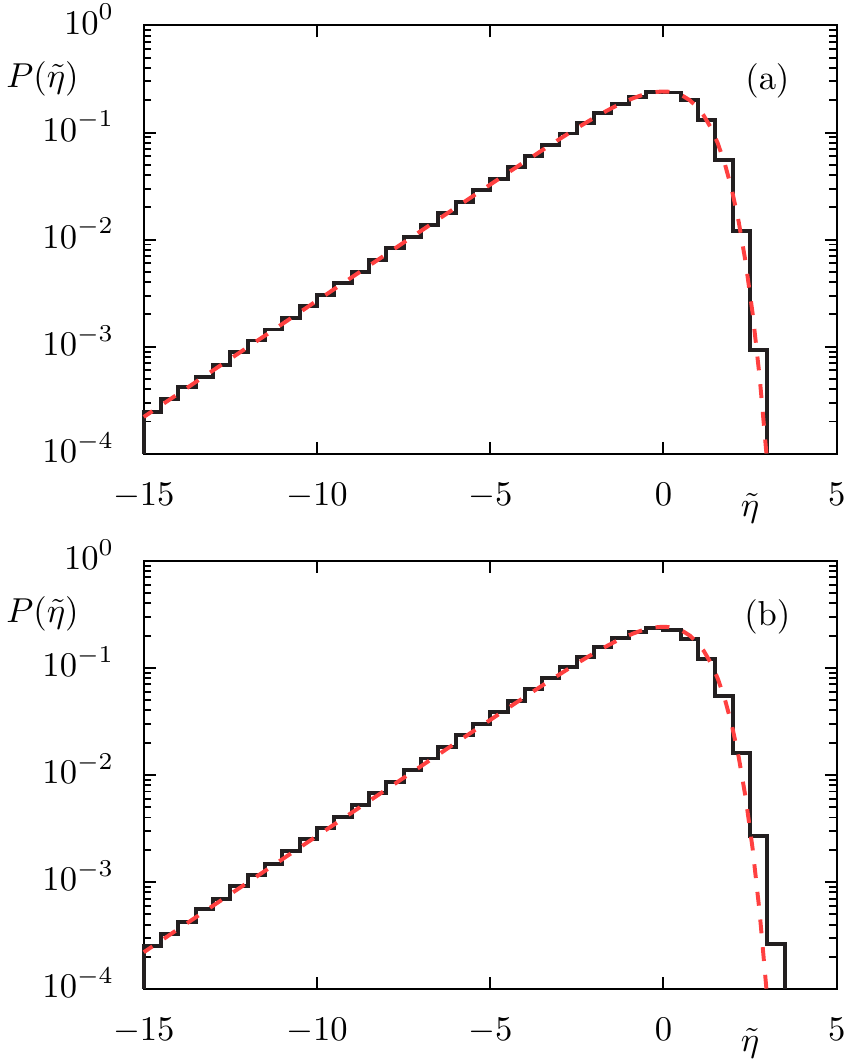}

        \caption{Distribution of $\tilde{\eta} = \ln(\eta)$ for
        the kicked spin chain with $L=12$.
        (a) shows the results for parameter set A and (b)
        for parameter set B, see table~\ref{tab:parameter_A_B}. The red
        dashed lines show the COE result~\eqref{eq:evec_stat_ln_COE}.}
        \label{fig:fig_evec_ln}
\end{figure}

In Fig.~\ref{fig:fig_spacing_evec_page} the eigenvector distribution $P(\eta)$
is shown in a logarithmic representation, where no differences are
visible between the chains~A and B.
To investigate the behaviour for small $\eta$
we consider the logarithm of the scaled and squared eigenvector
elements $\tilde{\eta} = \ln(\eta)$.
The logarithmic version of the Porter-Thomas
distribution~\eqref{eq:evec_stat_COE} reads
\begin{align}
        P_{\text{COE}}(\tilde{\eta})
                = \sqrt{\frac{\ue^{\tilde{\eta}}}{2\pi}}
                        \exp\left(-\frac{\ue^{\tilde{\eta}}}{2}\right)\;.
        \label{eq:evec_stat_ln_COE}
\end{align}
In Fig.~\ref{fig:fig_evec_ln} the distribution of $\tilde{\eta}$
for chain~A and chain~B is shown in a logarithmic representation. For small
values of $\eta$, corresponding to $\tilde{\eta} < 0$, we find for both chains
equally good agreement with the COE result.
Interestingly, we see here for large values of $\tilde{\eta}$, that the
distribution is a bit closer to the expectations for chain A than for chain B.
This is in line with the deviations we see for the eigenstate entanglement.
Thus, this can be seen as a hint that already the
distribution of the eigenvector components can show differences
from the expected behavior even if the level-spacing statistics agrees
well with the RMT results.

\section{Distribution of the coefficients}
\label{app:random_matrix_result}

Here we derive the distribution of the coefficients \eqref{eq:def_coeff}
of the effective spin interactions for
$H_{\text{eff}}^{\text{COE}}$ as defined in Eq.~\eqref{eq:H_eff_COE}
for random matrices, based on the known distributions of individual
eigenphases and eigenvector elements for COE matrices.

The effective Hamiltonian can be written in terms of the eigenphases
and eigenvectors of $U$ as
\begin{align}
 H_{\text{eff}} & = \ui \ln U \\
                & = \ui\, \mathbf{C}\mathbf{C}^{\dagger}\,
                        \ln U\,  \mathbf{C}\mathbf{C}^{\dagger}   \\
                & = \ui\, \mathbf{C}\,
                        \ln (\mathbf{C}^{\dagger}U\,  \mathbf{C})\,
                        \mathbf{C}^{\dagger}   \\
                & = - \mathbf{C}\, \text{diag}\left(\varphi_1,
                \dots,\varphi_N\right) \mathbf{C}^{\dagger}\;.
        \label{eq:H_eff_eigen}
\end{align}
Here we use the matrix
$\mathbf{C}$,
which contains the eigenvector elements,
$(\mathbf{C})_{nm} =c_m^{(n)} = \langle m|\psi_n\rangle$ for
the computational basis $\{|m\rangle\}_{m=1,\dots,N}$.
Thus, the coefficients can be computed by
\begin{align}
        C_{S_1, \dots, S_L} = -\frac{1}{N} \text{Tr}\big(&\left(S_1 \otimes
                \dots \otimes S_L \right) \nonumber \\
                &\times \mathbf{C} \text{diag}\left(\varphi_1,
                \dots,\varphi_N\right) \mathbf{C}^{\dagger}\big)\;.
\end{align}
The simplest non-trivial form of the dependence on
the phases and eigenvector elements of $U$ arises for the coefficient
\begin{align}
        C_{I,I,\dots, I,\sigma_z}
                = \frac{-1}{N} \sum_n \varphi_n
                        \sum_m (-1)^{m-1} |c^{(n)}_m|^2\;.
\end{align}
To find the distribution of the coefficient $C_{I,I,\dots, I,\sigma_z}$ we would have
to use the joint distribution of all eigenphases and eigenvector elements.
This distribution does not have a closed analytical from. Instead we assume
that the eigenphases and eigenvector elements are independent
of each other.
For the COE and CUE the eigenphases are
uniformly distributed in $[-\pi,\pi)$, i.e.\
\begin{align}
  P_{\text{COE/CUE}}(\varphi) = \frac{1}{2\pi}
  \quad\text{for} \quad
  \varphi \in [-\pi,\pi)\;.
\end{align}
This distribution has an average value $\overline{\varphi} = 0$
and finite variance $\sigma^2_{\varphi} = {\pi^2}/{3}$.
In the following, each eigenvector $|\psi_n\rangle$ is modeled using a
normalized state $|\phi_n\rangle$ with coefficients
$c_m^{(n)} = \langle m|\phi_n\rangle$.
In case of the COE the elements $c_m^{(n)}$
have to be real and
$\eta_{nm} = N |c_m^{(n)}|^2$ follows
the Porter-Thomas distribution
\begin{align}
        P_{\text{COE}}(\eta)
                = \frac{\exp(-\eta/2)}{\sqrt{2\pi\eta}}\;,
        \label{eq:evec_stat_COE_app}
\end{align}
which has average $\overline{\eta}_{\text{COE}} = 1$
and variance $\sigma^2_{\eta, \text{COE}} = 2$. For the CUE
$c_m^{(n)}$ can also become complex and
the distribution is an exponential,
\begin{align}
        P_{\text{CUE}}(\eta)
                = \ue^{-\eta}\;,
        \label{eq:evec_stat_CUE_app}
\end{align}
with average $\overline{\eta}_{\text{CUE}} = 1$
and variance $\sigma^2_{\eta, \text{CUE}} = 1$.

To derive the distribution of the coefficient
$C_{I,I,\dots, I,\sigma_z}$
we assume for simplicity that the matrix dimension $N$ is even
and first discuss the sum over $m$,
\begin{align*}
       \Sigma_m &:= \sum_m (-1)^{m-1} |c^{(n)}_m|^2\\
                &= \frac{1}{N} \sum_{m=1}^N (-1)^{m-1} \eta_{nm}\\
                &= \frac{1}{N}\left(
                        \sum_{m^{\prime}=1}^{N/2} \eta_{n,2m^{\prime}}
                        - \sum_{m^{\prime}=0}^{N/2} \eta_{n,2m^{\prime}+1}
                        \right)\\
                &= \frac{1}{2} \left(
                        \underbrace{\frac{1}{N/2}\sum_{m^{\prime}=1}^{N/2}
                                \eta_{n,2m^{\prime}}}_{I_1}
                        - \underbrace{\frac{1}{N/2}\sum_{m^{\prime}=0}^{N/2}
                                \eta_{n,2m^{\prime}+1}}_{I_2}\right)\;.
\end{align*}
The entries in both sums have the same distribution, given by
Eq.~\eqref{eq:evec_stat_COE_app} or
Eq.~\eqref{eq:evec_stat_CUE_app}, respectively.
Applying the central limit theorem for large $N$,
the distributions of $I_1$ and $I_2$ will both approach a normal distribution
with average zero and variance $\sigma^2_{I} = \sigma_{\eta}^2 / (N/2)$.
This implies that $\Sigma_m = 1/2 (I_1-I_2)$
is the scaled sum of two Gaussian random variables and thus itself
a Gaussian random variable with mean $\overline{\Sigma_m} = 0$ and variance
$\sigma^2_{\Sigma_m} = \frac{1}{2^2}(\sigma_I^2+\sigma_I^2) =
\sigma^2_{\eta}/N$, where
\begin{align}
        \sigma^2_{\Sigma_m, \text{COE}} = 2/N
        \;,\quad
        \sigma^2_{\Sigma_m, \text{CUE}} = 1/N\;.
\end{align}
We can now return to the distribution of the coefficient,
\begin{align}
        C_{I,I,\dots, I,\sigma_z}
                = \frac{-1}{N} \sum_n \varphi_n
                        \Sigma_m \;,
\end{align}
which can be interpreted as a product of two random
variables which follows the joint distribution of $\varphi$ and $\Sigma_m$.
Since the mean values $\overline{\varphi}$
and $\overline{\Sigma_m}$ are both zero also the mean of the joint
distribution is zero. The variance is given by
\begin{align*}
        \sigma^2_{\varphi\Sigma_m}
                = \sigma^2_{\varphi} \sigma^2_{\Sigma_m}
                + \sigma^2_{\varphi} \left(\overline{\Sigma_m}\right)^2
                + \sigma^2_{\Sigma_m} \left(\overline{\varphi}\right)^2
                = \frac{\pi^2}{3}  \sigma^2_{\Sigma_m}\;.
\end{align*}
Since this is a finite value we can apply the central limit theorem and
find that the coefficient $C_{I,I,\dots, I,\sigma_z}$
follow Gaussian distribution with mean value zero and variance
$\sigma^2 = \sigma^2_{\varphi\Sigma_m} / N$. Thus, we find
\begin{align*}
        \sigma^2_{\text{COE}} = \frac{\pi^2}{3} \frac{2}{N^2}
        \;,\quad
        \sigma^2_{\text{CUE}} = \frac{\pi^2}{3} \frac{1}{N^2}\;.
\end{align*}
As discussed before this distribution holds also for all the other
coefficients.

\end{document}